\documentclass[twocolumn,preprintnumbers,amsmah,amssymb,showpacs]{revtex4}
\usepackage{graphicx}
\usepackage{dcolumn}
\usepackage{bm}
 \usepackage[latin1]{inputenc}
\usepackage{multirow}
\usepackage{amsmath}
\usepackage{appendix}

\begin{document}                

\title{Neutron resonances in planar waveguides}

\author{S.V. Kozhevnikov$^{1*}$, V. K. Ignatovich$^{1}$, A. V. Petrenko$^{1}$,  and F. Radu$^2$}
\affiliation{$^{1}$Frank Laboratory of Neutron Physics, Joint Institute for Nuclear Research, 141980 Dubna, Russian Federation }
\affiliation{$^{2}$Helmholtz-Zentrum Berlin f\"ur Materialien und Energie, Albert-Einstein Strasse 15, D-12489, Berlin, Germany}
\date{\today}

\begin{abstract}Results of experimental investigations of a neutron resonances width in planar waveguides using the time-of-flight reflectometer REMUR of the IBR-2 pulsed reactor are reported and comparison with theoretical calculations is presented.  The intensity of the neutron microbeam emitted from the waveguide edge was registered as a function of the neutron wavelength and the incident beam angular divergence. The possible applications of this method for the investigations of layered nanostructures are discussed.     
\end{abstract}
\maketitle

\section{Introduction}

Neutron scattering is a powerful tool for investigations of magnetic and nonmagnetic solids, polymers and biological objects. The commonly used beams have the width of the order of 0.1 - 10 mm. For local investigation of samples the more narrow beams are required. The various focusing devices [1] are able to decrease the width of the neutron beam up to 50 {$\mu$}m. The smaller widths up to 0.1-10 {$\mu$}m can be achieved with the help of planar waveguides, which transform collimated neutron macrobeams into slightly divergent microbeams compressed in one dimension. The production of the unpolarized microbeam at the steady-state reactors was reported in [2,3] and the polarized one in [4]. The system of neutron microbeams at the time-of-flight reflectometer was registered in [5]. A combination of nonmagnetic neutron waveguide with polarized neutron reflectometer [6] was used for the investigations of an amorphous magnetic microwire [7,8] by the method of spin precession at transmission [9]. The application of the neutron microbeam for the investigation of another microstructure outside the waveguide we termed as Neutron Sonde Microscopy.

Inside a planar waveguide, the resonant enhancement of the neutron wavefunction density takes place and the enhanced neutron standing waves are formed. The planar waveguide is termed also as the resonator. The neutron standing waves method was developed in [10]. It is based on the theory of neutron resonances in layered structures described in [11]. The neutron standing waves can be registered as minima at the total reflection plateau or maxima of a characteristic radiation: gamma-quanta [12], alpha-particles [13], off-specular neutron scattering [14,15], neutron channelling [16] or spin-flip of polarized neutrons [17].

The phenomenon of the neutron wave propagation along the guiding layer inside the waveguide is termed as Neutron Channeling. The theory of neutron channeling in planar waveguides is described in [18]. The parameter of the neutron wavefunction density decay termed as the neutron channeling length was measured experimentally in [19,20].  In [21] we proposed the method of Polarized Neutron Channeling for the investigations of weakly magnetic thin films. In this case we register the neutron microbeam but the investigated system is a weakly magnetic layer inside a planar waveguide. The position of the microbeam peak on the neutron wavelength or the glancing angle of the incident beam corresponds to the waveguide structure and nuclear and magnetic potentials averaged over the guiding layer thickness. But the width of the microbeam peaks (or resonances) may deliver the information about dispersion of the nuclear or magnetic potentials.

In this communication we experimentally investigate the width of the resonances by registering the neutron microbeam emitted from the edge of a planar waveguide.

\section{Experimental setup}

\begin{figure}[ht]
       \includegraphics[clip=true,keepaspectratio=true,width=1\linewidth]{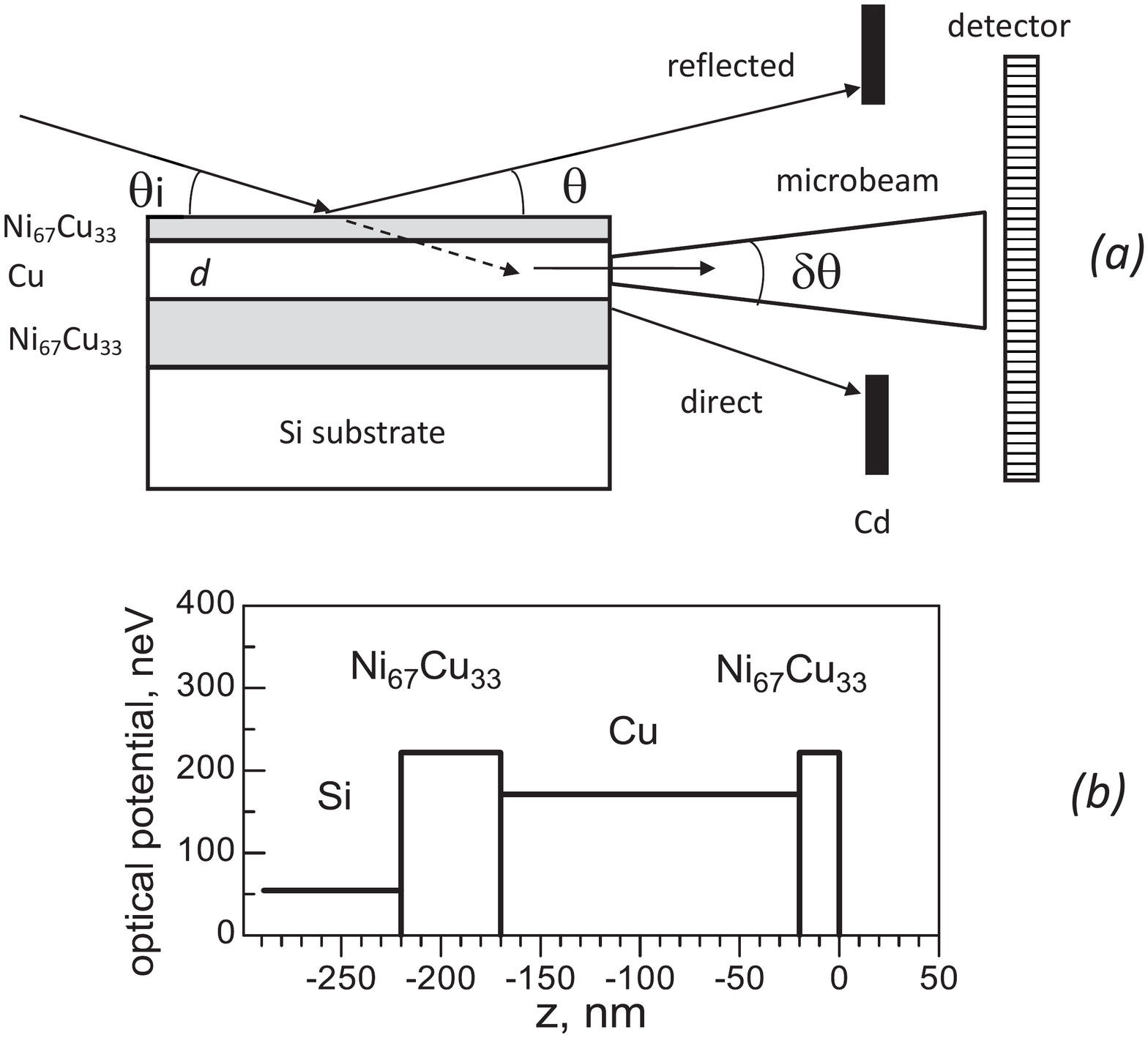}
          \caption{(a) Scheme of experiment. (b) Neutron optical potential of the sample as a function of the coordinate z perpendicular to the sample layers.}
   \label{fig1}
 \end{figure}

The experiment was done at the polarized neutron time-of-flight reflectometer REMUR [22] of the pulsed reactor IBR-2 (FLNP, JINR, Dubna, Russia). The scheme of the experiment is shown in Fig. 1a. The resonant planar waveguide is a multilayer Ni$_67$Cu$_33$(20 nm)/Cu(150)/Ni$_67$Cu$_33$(50)//Si(substrate) with optical nuclear potential shown in Fig. 1b. At room temperature the alloy Ni(67 \% at.)Cu(33 \% at.) is nonmagnetic. The initial neutron beam falls onto the sample surface under a grazing angle $\theta_i$, and tunnels through the upper layer into the wave guiding layer (channel) made of Cu. Inside the middle layer the neutron wave density at some wavelengths is resonantly enhanced. In the potential system Fig. 1b three resonances of orders n = 0, 1, 2 can be observed. The enhanced neutron wave after channeling along the guiding layer reaches the exit edge and leaks out through the gap of the width d equal to the thickness of the channeling layer. The outgoing microbeam has an angular divergence $\Delta \theta$, which is determined by the law of the Fraunhofer diffraction  $\Delta \theta \sim \lambda/d$, where $\lambda$ is the neutron wavelength. The divergent microbeam is registered by two-dimensional position-sensitive (PSD) $^3$He detector with spatial resolution 2 mm. The distance sample-detector was 4.94 m. The full time-of-flight base was 33.94 m. The glancing angle of the incident beam was fixed at 3.69 mrad. The angular divergence of the incident beam $\Delta \theta_i$ has been varied and the width of the microbeam intensity was measured as a function of the neutron wavelength.

\section{Experimental results}

The parameters of the sample found in a neutron reflectometry measurements were CuO(2.5 nm)/Ni$_67$Cu$_33$(14.9)/Cu(141.7)/Ni$_67$Cu$_33$(53.3)//Si(substrate). The nuclear potentials were found to be: CuO (45 neV), upper layer Ni$_67$Cu$_33$ (245 neV), Cu(171 neV), bottom layer Ni$_67$Cu$_33$ (219 neV), Si (54 neV).

\begin{figure}[ht]
       \includegraphics[clip=true,keepaspectratio=true,width=1\linewidth]{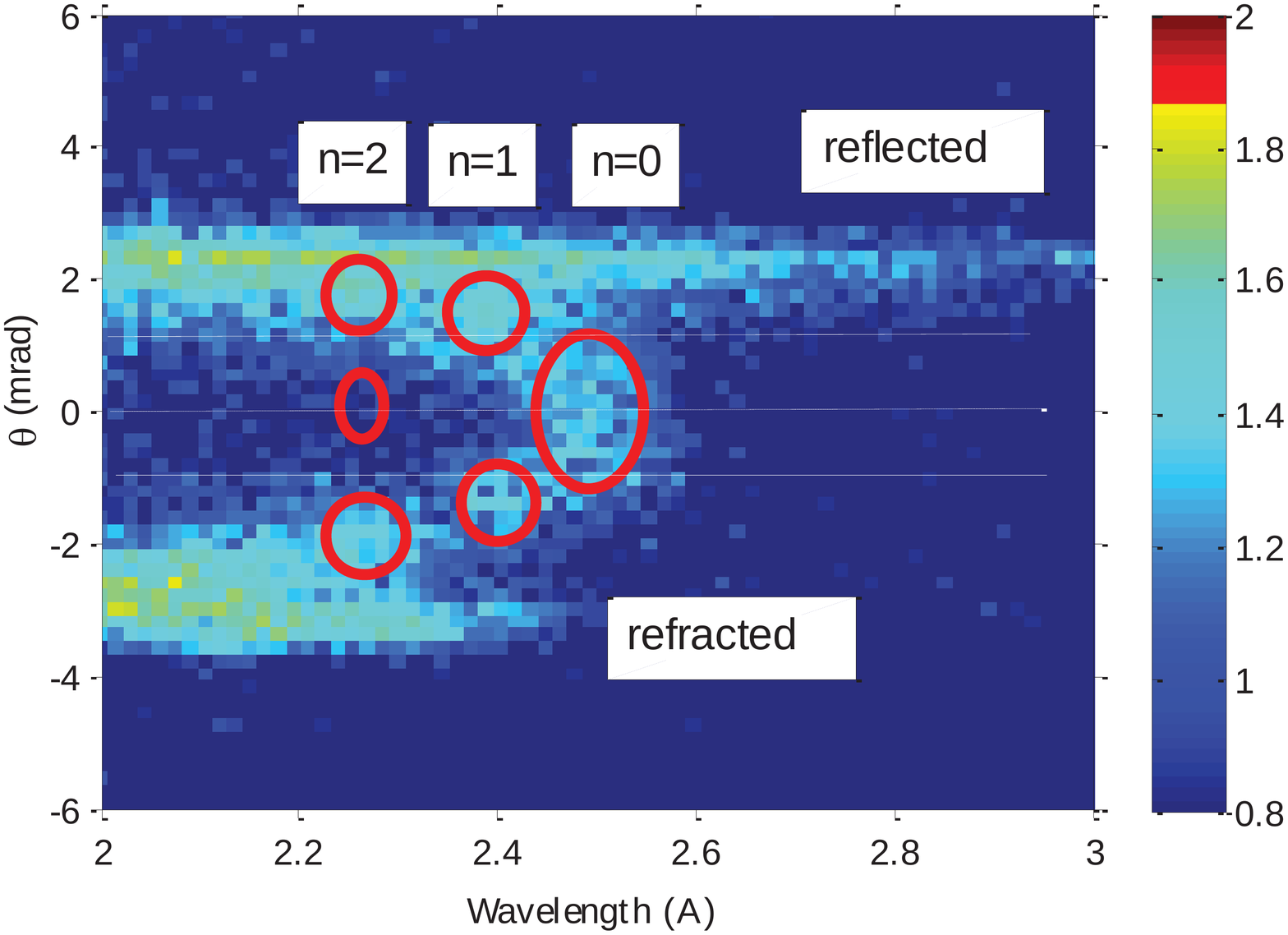}
          \caption{Two-dimensional map of the neutron intensity in dependence on the neutron wavelength and the scattering angle at the fixed grazing angle of the incident beam 3.69 mrad.}
   \label{fig2}
 \end{figure}

With time-of-flight technique it is possible to obtain at PSD a two-dimensional map of neutron counts for any glancing angle of the incident beam. One such a map for the glancing angle of the incident beam $\theta_i$ = 3.69 mrad is presented in Fig.2. The horizontal axis corresponds to neutron wavelength $\lambda$ and the vertical one corresponds to the outgoing angle $\theta$ with respect to horizon. The specularly reflected and direct beams were blocked by Cd beam-stops as shown in Fig. 1a. The counts near position of the direct beam correspond to the refracted one. The spots marked by ellipses correspond to the microbeams of different resonance orders n = 0, 1, 2. The space distribution of the neutron microbeam intensity reflects the two-dimensional distribution of the neutron wavefunction density $|\Psi_n(z,\lambda)|^2$  in the waveguide [3], where z is the coordinate perpendicular to the sample surface and n=0, 1, 2 ...  is the resonance order. 

\begin{figure}[ht]
       \includegraphics[clip=true,keepaspectratio=true,width=1\linewidth]{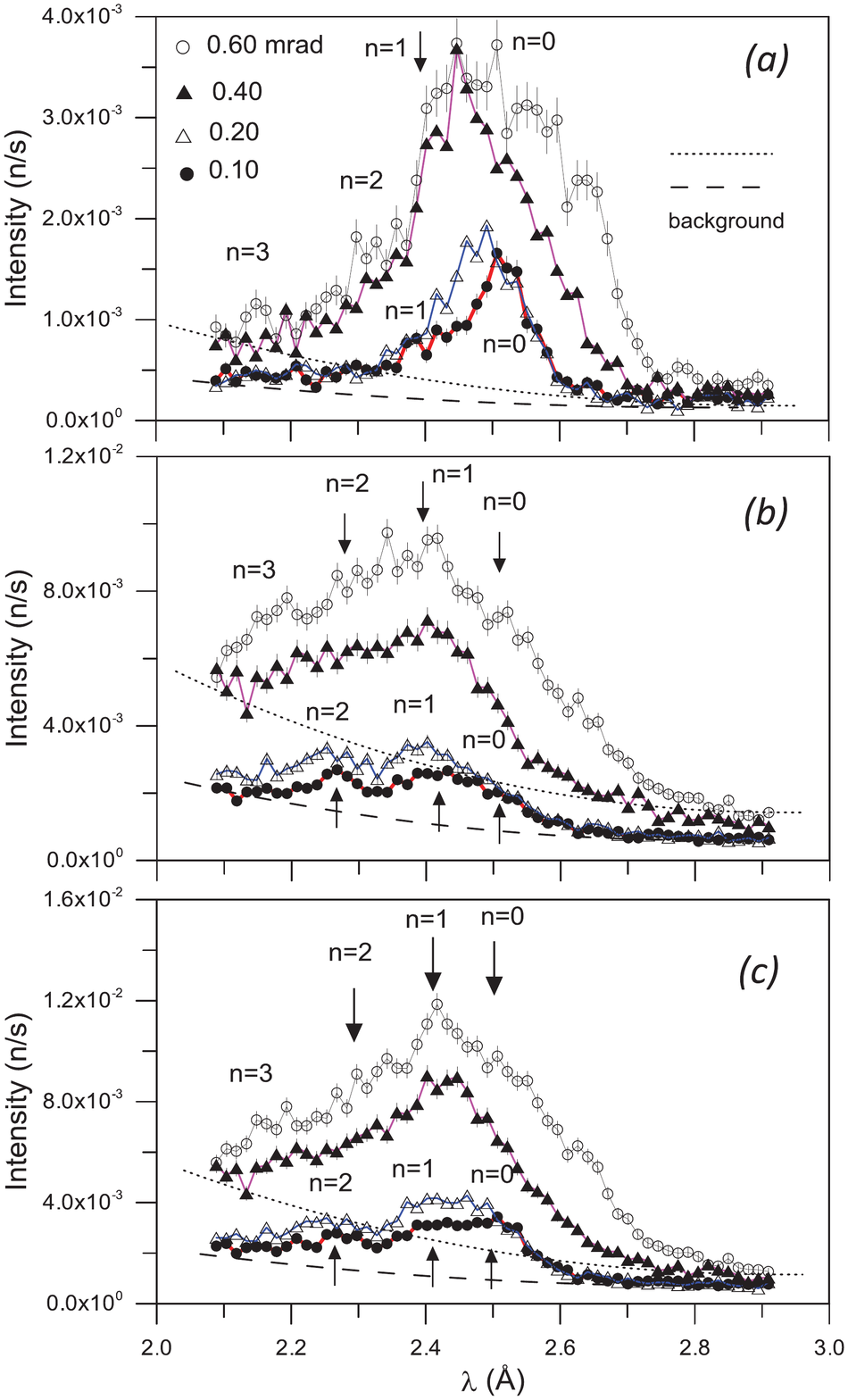}
          \caption{The microbeam intensity integrated around the sample horizon as a function of the neutron wavelength at various angular divergence of the incident beam: 0.60, 0.40, 0.20 and 0.10 mrad. The marked peaks correspond to resonances n=0, 1, 2, 3. (a) Narrow region near the sample horizon. (b) The region outside the sample horizon. (c) Wide region between the reflected and refracted beams.}
   \label{fig3}
 \end{figure}

In Fig. 3a, the dependence of the neutron intensity registered by PSD on the neutron wavelength is shown for the several angular divergences (FWHM) of the incident beam $\Delta\theta_i$ = 0.60, 0.4, 0,2 and 0.1 mrad. The intensity was integrated over the angular range between specularly reflected and the direct beams marked by two solid lines in Fig. 2. The strong peak is the resonance order n=0. The weak peak on the left side marked by the arrow is the resonance order n=1. It is the part of the peak n=1 because the intensity distribution of this resonance has a minimum near the horizon direction. One can see that the width of the peak n=0 decreases with decreasing the incident beam angular divergence. For the comparison, in Fig. 3b the neutron intensity integrated outside the central part near the horizon (Fig. 2) is presented. In this figures, the peaks of the resonances n=0 are absent and the peaks of the resonances n=1, 2, 3 are clearly seen. The neutron intensity in Fig. 3c is the sum of the integrated neutron intensities of Figs. 3a and 3b. 

In Fig. 4a the width (FWHM) of the resonance n=0 peak of Fig. 3a is shown as a function of the incident beam angular divergence (FWHM). Points are experimental ones and the solid line is a linear fit. One can see the linear decrease of the wavelength width with decrease of the angular divergence of the incident beam. The point cut-off by the line on the axis   mrad corresponds to the wavelength width $\Delta\lambda$=0.0356~$\AA$.

To estimate the eigen width of the resonance n=0, we have to extract the neutron wavelength resolution due to the reactor pulse width. In Fig. 4b the neutron count on the detector at the distance 33940 mm is presented as a function of the time (points are experiment ones and the line is a Gaussian fit). The fitted width of this peak (FWHM) is 280~$\pm$~6 ($\mu$s). This value of the reactor pulse width corresponds to the neutron wavelength resolution $\delta \lambda$=0.0326~$\pm$~0.0014~($\AA$) for the time-of-flight method. Thus, the experimental value of the internal width of the resonance n=0 can be estimated as $\Delta\lambda_{0,exp}$= ($\Delta\lambda-\delta\lambda$)=0.0030~$\pm$~0.0014 ($\AA$).                

\begin{figure}[ht]
       \includegraphics[clip=true,keepaspectratio=true,width=1\linewidth]{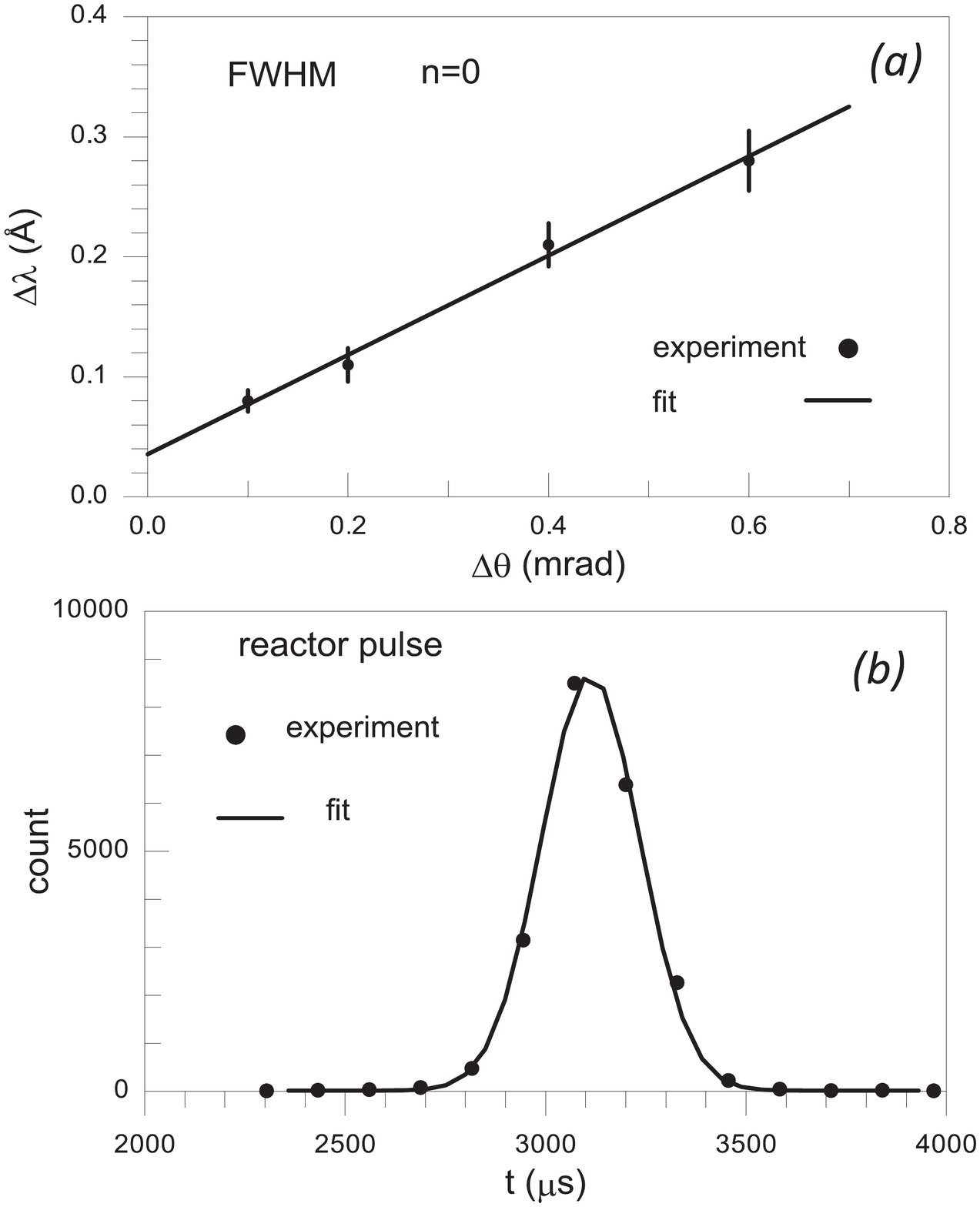}
          \caption{(a) The neutron wavelength width of the microbeam peak of the resonance n=0 measured experimentally shown as a function of the angular divergence of the incident beam (points). The line is the linear fit. (b) The fast neutron peak of the reactor pulse measured as a function of the time of the neutron arriving to the detector (symbols are experimental data and the line is the Gaussian fit). The peak position corresponds to the moment of the reactor burst and the peak width defines the neutron wavelength resolution. The right side of the experimental curve marked by the arrow contains the background.    }
   \label{fig4}
 \end{figure}

\section{Calculations}
Here we estimate the intrinsic width of the resonance peak n=0 using the theory of the resonances [11]. According to this theory the wave function in the resonant Cu layer can be represented as

 \begin{equation}
\begin{split}
\Psi(\bf r)= & exp({\it i} {\bf k_{||}} {\bf r_{||}}) \left [e^{\it i k_2 (z-d2)}+R_{32}e^{\it -i k_2(z-d_2)} \right] \\ & \frac{exp({\it i k_2 d_2}) {it T_{20}}}{1-exp(2{\it i k_2 d_2}) {\it R_{02} R_{32}}}\\
\end{split}
 \end{equation}
where $k_i=\sqrt{k^2_z-u_i}$  is the normal component of the neutron wave vector inside the potential $u_i$ of thickness $d_i$, $R_{ji}$, $T_{ji}$ are reflection and transmission amplitudes respectively from a potential $i$ to the potential $j$, and zero denotes vacuum. The vectors with index $||$ have components in the $(x,y)$ plane parallel to the sample surface.

\begin{figure}[ht]
       \includegraphics[clip=true,keepaspectratio=true,width=1\linewidth]{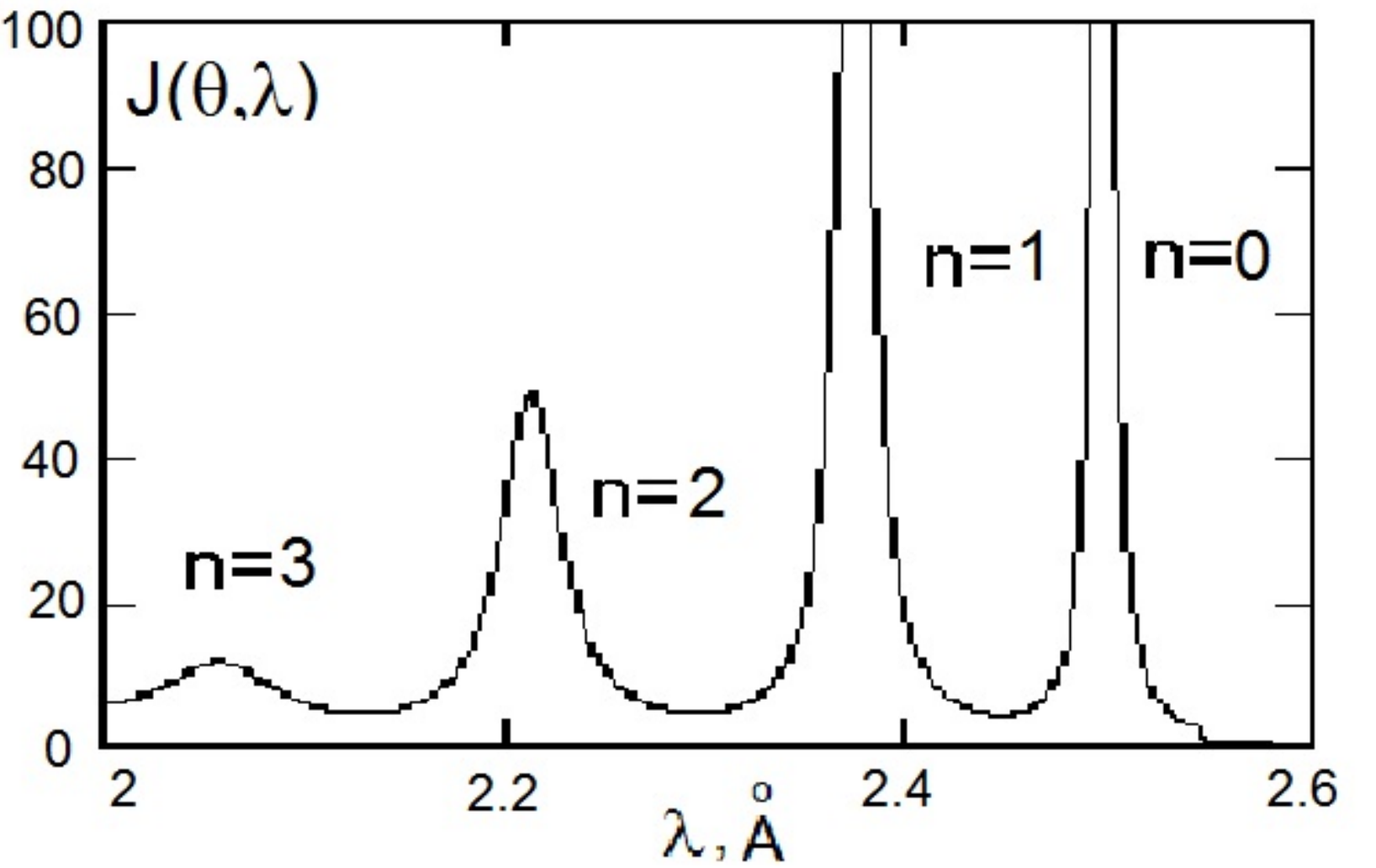}
          \caption{The microbeam intensity (11) integrated near the sample horizon as a function of the neutron wavelength at a given incidence glancing angle $\theta$=3.69 mrad.  Four resonances of orders n=0, 1, 2, 3 are seen. The calculated width $\Delta\lambda_0$ of the resonance n=0 is equal to 2.592$\times$10$^{-3}$~\AA. 
  }
   \label{fig5}
 \end{figure} 

The reflection amplitude from the right side of the full potential shown in Fig.1b can be represented in the form
 \begin{equation}
\begin{split}
R=  R_{20}+T_{20}T_{02}\frac{exp({\it i k_2 d_2} R_{32})}{1-exp(2{\it i k_2 d_2}) {\it R_{02} R_{32}}}
\end{split}
 \end{equation}
The amplitudes entering (1) and (2) are given by
 \begin{equation}
\begin{split}
&R_{20}=  R_{2}+\frac{T_2^2 r_{20}}{1-r_{20}R_2}, \qquad T_{20}=  \frac{T_2 (1+r_{20})}{1-r_{20}R_2}\\
&T_{20}=  \frac{T_2 (1-r_{20})}{1-r_{20}R_2}, \qquad R_{20}= -r_{20}+ \frac{(1-r_{20}^2)R_2}{1-r_{20}R_2}\\
&R_{32}=  -r_{20}+\frac{1-r_{20}^2)R_{43}}{1-r_{20}R_{43}}, \qquad R_{43}=R_3  \frac{T_3^2 r_{40}}{1-r_{40}R_3}
\end{split}
 \end{equation}
and in the last expression the Si substrate is supposed to be of infinite thickness. The amplitudes $R_i$ and $T_i$ are reflection and transmission amplitudes of a rectangular potential of height $u_i$ and thickness $d_i$:
 \begin{equation}
\begin{split}
&R_i=  r_{i0}\frac{1-exp({2\it i k_i d_i)}}{1-r_{i0}^2 exp(2{\it i k_i d_i})} \\ & T_i= exp({\it i k_i d_i})\frac{1-r_{i0}^2}{1-r_{i0}^2 exp(2{\it i k_i d_i})}
\end{split}
 \end{equation}
where $r_{i0}=(k_z-k_i)(k_z+k_i)$  is reflection amplitude from a potential step of height $u_i$. The denominator in (1) has resonant property. It becomes smallest $1-|R_{02}R_{32}|$ when in the guiding Cu layer the resonant phase condition 
 \begin{equation}
\begin{split}
&\gamma (\lambda)= 2k_2d +arg(R_{02})+arg(R_{32})=2\pi n
\end{split}
 \end{equation}
where integer n=0, 1, 2, ... denotes the order of the resonances. The resonances can be represented by the Breit-Wigner formula
 \begin{equation}
\begin{split}
\frac{1}{1-exp(i\gamma(\lambda))|R_{02}R_{32}|} & \approx   \frac{1}{1-|R_{02}R_{32}|[1+i\gamma^`(\lambda)(\lambda-\lambda_n)]}\\ 
& \approx  \frac{1}{\lambda-\lambda_n-i\Delta\lambda_n}
\end{split}
 \end{equation}
where 
 \begin{equation}
\begin{split}
\gamma^`(\lambda_n)=\frac{d}{dk}\gamma(k)_{\lambda=\lambda_n},
\end{split}
 \end{equation}
therefore the width of the resonance is 
 \begin{equation}
\begin{split}
\Delta \lambda=\frac{|R_{20}|R_{32}|-1}{|R_{20}|R_{32}|\gamma^`(\lambda_n)}
\end{split}
 \end{equation}
For the resonance n=0 the width is 
 \begin{equation}
\begin{split}
\Delta \lambda_0=2.592\times 10^{-3}\AA
\end{split}
 \end{equation}

The intensity outgoing the side edge of the channeling layer for the given incidence glancing angle $\theta$ and integrated over all outgoing angles can be represented in the form

 \begin{equation}
\begin{split}
J(\theta,\lambda)=C \int_0^{d_2}|\Psi({\bf r})|^2dz,
\end{split}
 \end{equation}
where C is some normalization constant, which for convenience is chosen here to be C=$10^{-3}$, and the wave function in the channel is given in (1). This intensity for a given $\theta$=3.69~mrad is shown in Fig. 5. Four resonances are clearly visible. Of course, their positions depend on the incidence angle. It is because that Eq. (6) contains $k_z$, which for thermal neutrons is $\approx \theta/\lambda$, where is the glancing angle of the incident beam, and its value for thermal neutrons is of the order of 10$^-3$. The larger is $\theta$, the larger is the resonance wavelength $\lambda_n$ in (7).

The experimentally measured resonance peaks presented in Fig. 3a can be theoretically calculated by the integration of (11) over the angular uncertainty $\Delta\theta$:

 \begin{equation}
\begin{split}
I(\lambda)= \int_\theta^{\theta+\Delta\theta}d\theta`J(\theta`,\lambda).
\end{split}
 \end{equation}

\begin{figure}[ht]
       \includegraphics[clip=true,keepaspectratio=true,width=1\linewidth]{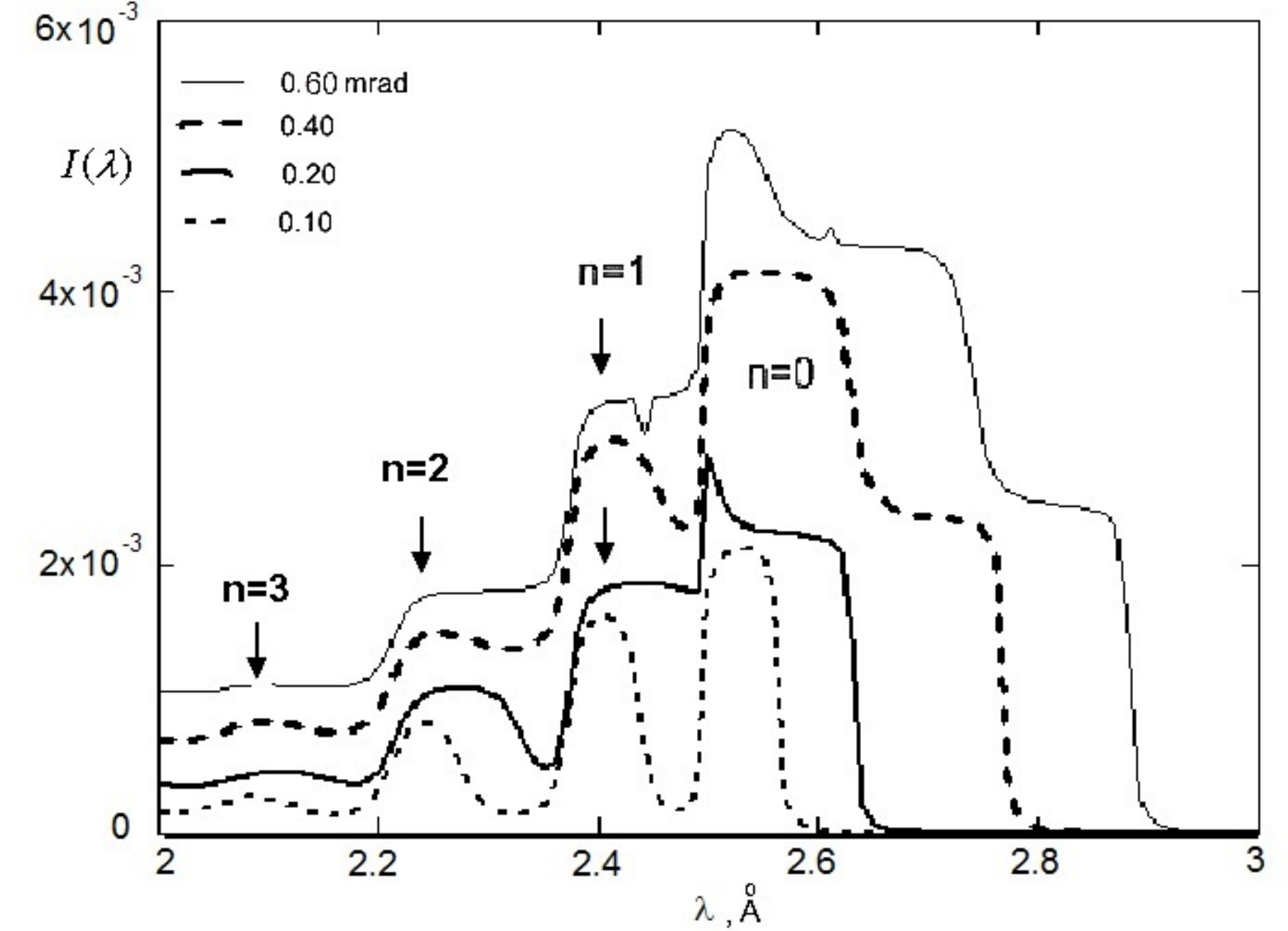}
          \caption{The microbeam intensity (11) integrated near the sample horizon and integrated over different angular uncertainties $\Delta\theta$ of the incident beam near $\theta$=3.69 mrad as a function of the neutron wavelength. Four resonances of orders n=0, 1, 2, 3 are seen.  The curves qualitatively agree with experimental ones shown in Fig. 3a. }
   \label{fig6}
 \end{figure}
The result is presented in Fig. 6. The calculated peaks are in good agreement with experimental ones shown in Fig. 3a. It is also clear that the width of the n-th peak, $\Gamma_{\lambda n}$, can be represented as a linear function

 \begin{equation}
\begin{split}
\Gamma_{\lambda n}\approx \Delta\theta\lambda_n+\Delta\lambda_n ,
\end{split}
 \end{equation}
which decreases with decrease of the angular uncertainty and in the limit $\Delta\theta$=0 gives $\Delta\lambda_n$ (9) .

One can see that value of the calculated width (9) of the resonance n=0 coincides with the experimentally estimated value within error bar due to the reactor pulse width.  

\section{Discussion}
Neutron resonances in planar waveguides can be registered in two ways. One method is to observe minima on the total reflection plateau of the specular reflectivity. Another way is to observe maxima of the secondary radiation such as alpha-particles and gamma-rays or the secondary channel of the neutron radiation like off-specular neutron scattering, spin-flip and the neutron channeling. The secondary characteristic radiation can be used only with special matter like $^6$Li or Gd. The secondary channel of neutron radiation can be used for many materials: magnetic and nonmagnetic ones. 

The neutron interaction with matter is weak as a rule. Therefore the dips on the total reflection plateau at resonances in planar waveguides are not deep (see for instance [14,15]).  At some special conditions the dips at resonances can reach 0.5 of the total reflection [12,13]. It means that in the best case the ratio \textit{effect/background}=0.5. The typical value is \textit{effect/background}=0.1.

In the case of the neutron channeling and the microbeam registration, the ratio effect/background can be 10 as one can see in Fig. 3a. It is 100 times better. In the microbeam geometry one can effectively separate the useful neutrons from the background ones contained in reflected, refracted and direct beams. It gives the possibility to observe a low effect at almost zero level of the background.           

\section{Conclusion}

The width of the neutron resonances in planar waveguides was investigated experimentally for the first time using the neutron channeling and the registration of the microbeam emitted from the sample edge. Time-of-flight technique was used to measure the microbeam peak width in dependence on the neutron wavelength and the incident beam angular divergence. The linear dependence was found. The measured experimental resonance peaks are well reproduced in calculations, which proves that mathematical model for description of resonant channeling is quite correct.  We expect that method of resonant channeling described in this communication is quite sensitive and can be applied for investigation of small effects of nuclear or magnetic inhomogeneities of the matter in layered nanostructures.


\end{document}